\documentclass{article}
\usepackage{spconf,amsmath,graphicx}
\usepackage{amssymb}
\usepackage{amsthm}
\usepackage[applemac]{inputenc}

\newcommand {\R} {{\mathbb{R}}}

\newcommand {\E} {{\mathbb{E}}}
\newcommand {\om} {{\omega}}
\newcommand {\la} {{\lambda}}
\newcommand {\La} {{\Lambda}}
\newcommand{\hS}{\widehat S}

\newcommand {\oS} {{\overline S}}

\title{Audio Texture Synthesis with Scattering Moments}
\twoauthors
  {Joan Bruna}
   {New York University\\
   Courant Institute\\
   New York, NY}
  {Stéphane Mallat
\thanks{This work is supported by the ANR 10-BLAN-0126 and ERC InvariantClass 320959 grants.}}
   {Ecole Noramle Sup\'erieure\\
  Computer Science Department\\
   Paris, France}
\begin{document}
%
\maketitle
\begin{abstract}
We introduce an audio texture synthesis algorithm based on scattering moments.
A scattering transform is computed by iteratively decomposing a signal
with complex wavelet filter banks and computing their amplitude envelop.
Scattering moments 
provide general representations of stationary processes
computed as expected values of scattering coefficients. They are estimated
with low variance estimators from single realizations. Audio signals having
prescribed scattering moments are synthesized with a gradient 
descent algorithms. Audio synthesis examples show that scattering representation provide good synthesis of audio textures with much fewer coefficients than the state of the art.
\end{abstract}

\section{Introduction}

The representation of a non-Gaussian stationary process 
remains a fundamental
issue of probability and statistics. Signal processing faces many such issues,
in particular for auditory and image textures, which can be modeled
as realizations of highly non-Gaussian processes. A random vector
$X \in \mathbb{R}^N$ can be represented by a vector of generalized moments
$\Phi X = \{ \E(\phi_n (X)\}_n$ which project the distribution of $X$
over multiple functions $\phi_n (x)$ with $x \in \R^N$. 
Random signal synthesis can then 
be performed by sampling the maximum entropy
distribution, which is a Boltzmann distribution 
whose generalized moments are specified by $\Phi X$. 
For most signal
processing applications, one needs to estimate $\E(\phi_n (X)$ from
a single realization of $X$, by replacing the expected value with a
spatial or time average. We concentrate on
on audio texture synthesis, which is an important application.
The information loss of the representation can be checked
by evaluating the perceptual quality of synthesized signals.

Second order moments 
lose essential perceptual information in audio and image signals because
they provide no information on non-Gaussian behavior. 
High order moments are rarely used
because their estimation from a single realization
has a variance which is too large.
Representations based on generalized moments have been proposed to
represent and synthesize audio and image textures, often based on histograms
of non-linear transformations of the signal
\cite{simoncelli, audiohos}. 
Simoncelli and McDermott 
have obtained particularly efficient results from covariance measurements
at the output of multistage filter banks \cite{simoncelli_mcdermott}. In the following
we propose an audio texture representation and a synthesis algorithm based
on scattering moments. 

Scattering transforms have recently been introduced
\cite{stephane,pami,deepscatt,laurent} to represent 
audio signals and images, 
while providing state of the art results for texture
discrimination, and genre recognition in audio \cite{deepscatt}.
A scattering transform iterates 
on complex wavelet transforms and modulus operators which
compute their envelop. It has close relations with 
psychophysical and physiological models 
\cite{shamma,dau,slaney}.
For stationary processes, it estimates
a vector of expected values called scattering moments.
This paper shows that scattering moments
provide a compact representation of stationary processes, which encodes
important non-Gaussian properties 
arising from multiscale amplitude and frequency modulations. 
This is demonstrated through audio synthesis.

Section \ref{scattsect} reviews the properties of scattering moments
for auditory signals. An efficient audio synthesis algorithm is
described in Section \ref{algosect}.
Section \ref{numerical} gives synthesis results on natural audio textures.
Computations can be reproduced with a software available at 
{\it www.di.ens.fr/data/software/scatnet}.

{\it Notations:} $\widehat{x}(\om)=\int x(t) \exp(-i \om t) dt$ is 
the Fourier transform of $x(t)$. We denote 
$\E(X )$ the expected value of a stationary process $X(t)$ at any $t$,
and $\sigma^2(X) = E(|X|^2) - E(X)^2$.

\section{Scattering Moments}
\label{scattsect}

A scattering transform characterizes transient structures through
high order coefficients which capture modulation properties.
They are computed by iterating on filter banks
of complex wavelet filters.

\subsection{Wavelet Filter Bank}

A wavelet $\psi (t)$ is a band-pass filter. We
consider a complex wavelet with a quadrature phase, 
whose Fourier transform satisfies
$\widehat \psi(\om) \approx 0$ for $\om < 0$.
We assume that the center frequency of $\widehat \psi$ is $1$ and 
that its bandwidth is of the order of $Q^{-1}$. 
Wavelet filters centered
at the frequencies $\lambda = 2^{j/Q}$ are computed by dilating $\psi$:
\begin{equation}
\label{psidilation}
\psi_\la (t) = \la\, \psi(\la\, t)~~\mbox{and hence}~~
\widehat \psi_\la (\om) = \widehat \psi(\la^{-1} \omega)~.
\end{equation}
We denote by $\Lambda$ the index set of $\la = 2^{j/Q}$ over
the signal frequency support, 
and we impose that these filters fully cover the positive frequencies
\begin{equation}
\label{consdf}
\forall \om>0~,~ 1- \epsilon 
\leq \frac 1 2 \sum_{\la \in \Lambda} |\widehat{\psi}_{\lambda}(\omega)|^2 \leq 1~.
\end{equation}
for some $\epsilon <1$.
The wavelet transform of a random process $X(t)$ is
\[
W X = \{X \star \psi_\la(t)  \}_{\la \in \Lambda}~.
\]
One can derive from (\ref{consdf}) that the variance
satisfies
\begin{equation}
\label{lp_basic}
\sigma^2( X) (1- \epsilon) \leq \sum_{\la \in \Lambda} \E(|X \star \psi_\la|^2) \leq \sigma^2(X)~.
\end{equation}

\subsection{Scattering Moments}

Scattering moments provide a representation of stationary processes,
with expected values of a non-linear operator, calculated by
iterating over wavelet transforms and a modulus. 
First order scattering coefficients
are first order moments of wavelet coefficient amplitudes:
\[
\forall la_1 \in \La~~,~~S X(\la) = \E(|X \star \psi_\la|)~. 
\]
The Q-factor $Q_1$ adjusts the frequency resolution of 
these wavelets. First order scattering moments
provide no information on the time-variation of the scalogram
$|X \star \psi_{\la_1} (t)|$. 
It averages all audio modulations and 
transient events, and thus lose perceptually important information.
 
Second order scattering moments recover information on
audio-modulations and transients by computing
the wavelet coefficients of each $|X \star \psi_{\la_1}|$, and their
first order moment:
\[
\forall \la_2~~,~~
S X(\la_1,\la_2) = \E(||X \star \psi_{\la_1}| \star \psi_{\la_2}|)~ .
\]
These multiscale variations of each envelop $|X \star \psi_{j_1}|$,
specify the amplitude modulations of $X(t)$ \cite{deepscatt}.
The second family of wavelets $\psi_{j_2}$ typically have
a $Q$-factor $Q_2=1$ to accurately measure the sharp transitions of
amplitude modulations.
Scattering coefficients have a negligible amplitude for
$\la_2 > \la_1$ because $|X \star \psi_{\la_1}|$ is then a regular envelop
whose frequency support is below $\la_2$. Scattering coefficients are thus
computed only for $\la_2 < \la_1$. 

Applying more wavelet transform envelops defines
scattering moments at any order $m \geq 1$:
\begin{equation}
\label{expansdf}
\oS X(\la_1,...,\la_m) =  \E(|~|X \star \psi_{\la_1}| \star ... | \star \psi_{\la_m}|)~.
\end{equation}
By iterating on the inequality (\ref{lp_basic}), one can 
verify \cite{stephane}
that the Euclidean norm of scattering moments
\begin{equation}
\label{enfsondf}
\|\oS X\|^2 = \sum_{m=1}^{\infty} \sum_{(\la_1,...,\la_m) \in \La_m}
|\oS X (\la_1,...,\la_m)|^2 .
\end{equation}
satisfies
\[
\|\oS X\|^2 \leq  \sigma^2(X)~.
\]
Expected scattering coefficients are first moments of 
non-linear functions $X$ and thus depend upon high order moments
of $X$ \cite{stephane}. But as opposed to high order moments,
the scattering representation is computed with wavelet transforms and
modulus operators, which do not amplify the variability of $X$.
It results into low-variance estimators.

Scattering moments are estimated 
by replacing the expectation with a time averaging over the signal
support. Suppose that $X(t)$ is defined for $0 \leq t < N$. With periodic
border extensions, we compute empirical averages
\begin{equation}
\label{scanesf}
\hS X(\la_1,...,\la_m) = N^{-1} \sum_{t=1}^N
|~|X \star \psi_{\la_1}| \star ... | \star \psi_{\la_m}(t)|~.
\end{equation}

For most audio textures, the energy of the scattering
vector $\|\oS X\|^2$ is concentrated over first and second order
moments \cite{deepscatt}. We thus only compute $\hS X(\la_1)$ and
$\hS X(\la_1,\la_2)$ for $1 \leq \la_1 = 2^{j_1/Q_1} \leq N$
and $1 \leq \la_2 = 2^{j_2/Q_2} < \la_1$. 
Scattering moments estimators  have large variance at the
lowest frequencies because the wavelet coefficient amplitudes
are highly correlated in time. These higher variance estimators are
removed by keeping only the frequencies $\la_1$ and $\la_2$ above
a fixed frequency $N_0$. We thus compute
$Q_1 \log_2 (N/N_0)$ first order scattering moments and
$Q_1 Q_2 (\log_2 N/N_0)^2/2$ second order scattering moments.

Scattering transforms have been extended along the frequency
variables to capture frequency variability and provide transposition
invariant representations \cite{deepscatt}. 
Transpositions refer to translations
along a log frequency variable. 
For audio synthesis, this frequency transformation
will only be performed on first order coefficients.
We denote $\gamma = \log_2 \lambda_1$, and
define 
wavelets $\bar \psi_{\bar \la} (\gamma)$ having an octave bandwidth of $Q = 1$.
The corresponding wavelet transform is thus computed with convolutions
along the log-frequency variable $\gamma$.

The scalogram is now considered as a function of $\gamma$ for each fixed 
time $t$:
\[
F_t (\gamma) = |X \star \psi_{2^{\gamma}} (t)|~.
\]
Second order frequency scattering moments are the first
order moments of the wavelet coefficients of $F_t (\gamma)$ computed
along $\gamma$:
\[
\oS X (\la_1, \bar \la_2) = 
\E (|F_t \star \bar \psi_{\bar \la_2} (\log_2 \la_1)|)~,
\]
This expected value is estimated with a time averaging
\begin{equation}
\label{time-fnsdf}
\hS X (\la_1, \bar \la_2) = 
N^{-1} \sum_{t=1}^N |F_t \star \bar \psi_{\bar \la_2} (\log_2 \la_1)|~.
\end{equation}
If $K=Q_1 \log_2 (N/N_0)$ is the total number of first order 
scattering moments, the number of second order frequency
scattering coefficients is $ \alpha K$, where $\alpha$ is 
an oversampling constant which is set to $2$ in our experiments. 

\section{Scattering Synthesis}
\label{algosect}
We present a gradient descent algorithm 
on the scattering domain to adjust
 scattering moments estimated from available 
 observations.
 
A maximum entropy distribution satisfying  
a set of moment conditions is a Gibbs distribution
defined by the Boltzmann theorem.
Sampling this distribution is possible
with the Metropolis-Hastings algorithms but it is computationally very
expansive in high dimension. This algorithm is often approximated 
with a gradient descent algorithm. It is initialized with 
a Gaussian white noise realization, whose moments are progressively
adjusted by the gradient descent 
\cite{zhu_et_al,simoncelli, simoncelli_mcdermott}.

Let $Y(t)$ be the realization of an 
auditory texture of $N$ samples. A vector of first order and second
order scattering moment estimators
$\hS X$ is computed with (\ref{scanesf}). 
This vector may also include second order frequency
scattering moments (\ref{time-fnsdf}).
To synthesize a
new audio signal $X$ such that $\hS X = \hS Y$,
we start with a realization of white Gaussian noise $X_0$.  
At each iteration $n$, we want to  minimize
\begin{equation}
\label{ener0}
E(X) = \frac 1 2  \| \hS X_n - \hS Y \|^2  .
\end{equation}
A gradient descent computes
\begin{equation}
\label{gradesc}
X_{n+1} = X_{n} -  \gamma \nabla E(X_{n}) = X_{n} -  \gamma \partial \hS X_n^T ( \hS X_n - \hS Y)~,
\end{equation}
where $\partial \hS X_n$ is the Jacobian of $\hS X$ with respect to $X$, evaluated at $X_n$, 
and $\gamma$ is a gradient step, which is kept fixed 
at a sufficiently small value for the sake of simplicity.

The minimization of (\ref{ener0}) is a non-linear least squares problem. 
The Levenberg-Marquardt Algorithm (LMA) \cite{numeropt} 
significantly accelerates the convergence. 
It replaces $\partial \hS X_n^T$ in (\ref{gradesc})
by the pseudoinverse
\[
\partial \hS X_n^\dagger = (\partial \hS X_n^T \partial \hS X_n)^{-1} \partial \hS X_n^T,
\] 
which requires computing a pseudoinverse on each iteration. 
The LMA typically requires $20$ iterations to reach a relative
approximation error of $10^{-2}$ and $40$ to reach $10^{-4}$,
tested on the collection of auditory textures described in next section.

%

\section{Numerical Experiments}
\label{numerical}

The audio scattering synthesis algorithm is tested on a 
dataset of natural sound textures of McDermott and Simoncelli, 
available at \cite{mcdermott_website}. It is
a collection of $15$ sound textures, of $7$ seconds
each, sampled at $20$ KHz, thus including $N\sim 10^5$ samples.
Our synthesis results are available at \cite{scatlink}.

McDermott and Simoncelli \cite{simoncelli_mcdermott} have
constructed an audio representation based on physiological models
of audition. Similarly to a scattering transform,
it uses two constant-Q filter banks. 
The first set of \emph{cochlea} filters  
consists in $30$ complex bandpass filters.
Their envelop is first compressed with a 
contractive nonlinearity and then redecomposed with a new filter
bank. They extract a collection 
of $1500$ coefficients, comprising marginal moments of each 
cochlea envelop and their corresponding modulation bands, 
as well as pairwise cross-correlations across different cochlea
and modulation bands. In \cite{torsten}, the authors used
a similar model to produce a texture representation with about
$800$ coefficients. 

\begin{figure*}
  \centering
  \centerline{\includegraphics[width=18.0cm,trim=0 0 0 0.2in,clip]{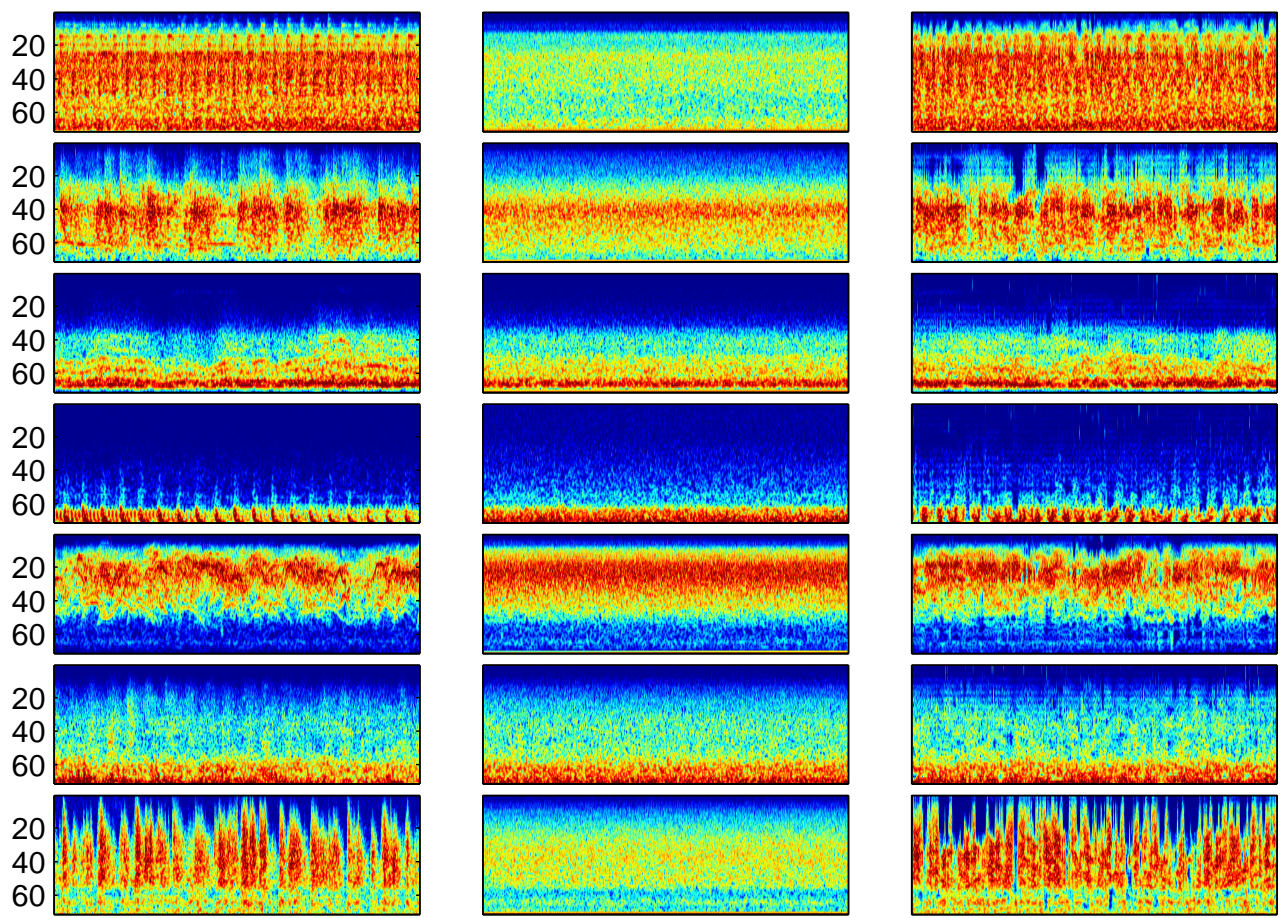}}
%
\caption{ Each image the scalogram of an audio recording: 
time along the horizontal axis and log-frequency 
up to $10$KHz along the vertical axis.
Left column:  original audio textures from \cite{mcdermott_website}. 
Middle column: Reconstruction from 1st order time scattering moments
Right column: reconstruction from 1st and 2nd order
time scattering moments.
The sounds are produced (from to to bottom) by
jackhammer, applause, wind, helicopter, sparrows, 
train, rusting paper. 
}
\label{texture_synthesis}
\end{figure*}

Scattering audio synthesis is performed with much fewer coefficients.
With $Q_1 = 4$ and $N_0=2^2$ 
there are
$Q_1 \log_2 N/N_0 = 46$ first order moments, 
$Q_1 Q_2 (\log_2 N/N_0)^2/2 = 266$ second order moments
and $2 \cdot 46=92$ frequency scattering moments
The total representation thus has $402$ coefficients.
Figure \ref{texture_synthesis} shows the scalogram of signals
recovered from first order moments only or first and second
order moments.
Reconstructions from first order moments are essentially
realizations of Gaussian processes. They do not capture the
transient and impulsive structures of the textures, such 
as the hammer or the applause.
When second order scattering moments are included, 
the reconstructed textures contain these 
highly non-Gaussian phenomena, which produce
highly realistic synthesized sounds. 
Scattering moments have the ability to capture 
processes with irregular spectra, such as the jackhammer, 
as well as wideband phenomena such as fire cracking 
or applause.

\begin{figure*}
  \centering
  \centerline{\includegraphics[width=8.5cm,trim=0.7in 0.4in 0.2in 0.1in,clip]{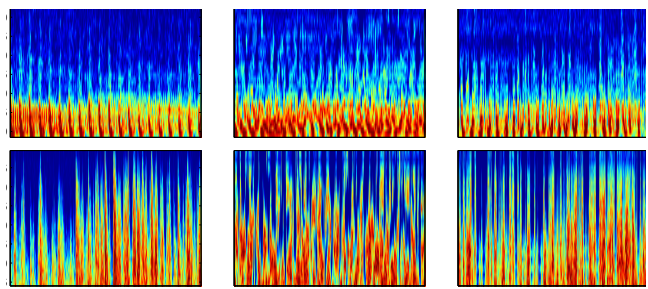}}
\hfill
\caption{ 
Impact of frequency scattering moments. 
Left column: original signals. Middle column: synthesis from
first and second order time scattering moments.
Right column: synthesis obtained by adding frequency 
scattering moments.
Observe how without frequency scattering, the subbands
tend to decorrelate, which prevents synthesizing 
impulsive phenomena.
The sounds are produced by a 
helicopter and rusting paper. More examples 
available at {\it cims.nyu.edu/$\sim$bruna}.
}
\label{freqscatt_comp}
\end{figure*}

Figure \ref{freqscatt_comp} shows that
frequency scattering moments correlate and thus synchronize the
amplitude variations across frequency bands. This is necessary
to accurately reproduce transient structures in textures.
The synthesis of wide-band textures can be further improved
by combining scattering moments computed with dyadic
wavelets having $Q_1=1$. It adds $120$ coefficients which further
 constraint the frequency interferences created by time varying
 modulations.

\section{Conclusions}

A texture audio synthesis is performed with a gradient
descent algorithm which progressively 
adjusts the scattering moments of a signal. 
Good perceptual reconstructions are obtained 
with fewer coefficients than state of the art algorithms.

First and second order scattering moments are thus efficient texture descriptors; 
on the one hand, they are sufficiently informative so that realizations with similar
coefficients have good perceptual similarity. On the other hand, they are 
 consistent: realizations of the same process (hence perceptually similar)
have similar scattering representations, as opposed to high order  
moments.

\pagebreak

\bibliographystyle{IEEEbib}
\bibliography{strings,refs}

\end{document}